# Kepler observations of rapid optical variability in the BL Lac object W2R1926+42

R. Edelson [1,], R. Mushotzky [1, 2], S. Vaughan [3], J. Scargle [4],
P. Gandhi [5], M. Malkan [6], W. Baumgartner [2]

## ABSTRACT

We present the first Kepler monitoring of a strongly variable BL Lac, W2R1926+42. The light curve covers 181 days with ~0.2% errors, 30 minute sampling and >90% duty cycle, showing numerous $\delta I/I > 25\%$ flares over timescales as short as a day. The flux distribution is highly skewed and non-Gaussian. The variability shows a strong rms-flux correlation with the clearest evidence to date for non-linearity in this relation. We introduce a method to measure periodograms from the discrete autocorrelation function, an approach that may be well-suited to a wide range of Kepler data. The periodogram is not consistent with a simple power-law, but shows a flattening at frequencies below $7 \times 10^{-5}$ Hz. Simple models of the power spectrum, such as a broken power law, do not produce acceptable fits, indicating that the Kepler blazar light curve requires more sophisticated mathematical and physical descriptions than currently in use.

*Subject keywords:* BL Lacertae objects: general – BL Lacertae objects: individual (W2R1926+42) – galaxies: active

## 1. Introduction

BL Lacertae objects are defined by their rapidly-variable, featureless and polarized γ-ray to radio continua. They are compact sources often exhibiting apparent "superluminal" motion. These properties motivated development of models in which the vast bulk of luminosity is due to Doppler-boosted radiation from a relativistic jet (e.g., Scheuer & Readhead 1979, Blandford & Konigl 1979). The variability is thought to arise from shocks propagating down the jet (e.g., Marscher & Gear 1985, Marscher et al. 2008).

It is in the optical that the term BL Lac "microvariability" (detectable variability within a single night) was first used. Much of the early observational effort (e.g., Carini et al. 1990, Wagner 1991) was devoted to establishing that this phenomenon is real, and not a result of systematic errors due to e.g. the atmospheric seeing variations that limit ground-based errors to ~1%. However problems inherent in ground-based optical monitoring, in particular, the difficulty of obtaining continuous coverage over more than a single ~12 hr night, have limited further observational progress.

The Kepler satellite (Borucki et al. 2010) overcomes these limitations with order-of-magnitude improvements in precision, sampling, and duty cycle over the best currently available from the ground. This paper reports the first Kepler monitoring of a highly variable BL Lac object, W2R1926+42, allowing analyses that have previously been impossible for this class of objects. This paper is organized as follows: Section 2 presents the Kepler observations and data reduction, Section 3 gives the results of various time-series analyses applied to these data, Section 4 discusses the theoretical implications of these results and Section 5 gives some concluding remarks.

[1] Department of Astronomy, University of Maryland, College Park, MD 20742-2421, USA
[2] Laboratory for High Energy Astrophysics, NASA/GSFC, Code 662, Greenbelt, MD 20771, USA
[3] University of Leicester, X-ray and Observational Astronomy Group, Department of Physics and Astronomy, University Road, Leicester, LE1 7RH, United Kingdom
[4] Astrobiology and Space Science Division, NASA Ames Research Center, Moffet Field, CA 94035, USA
[5] Institute of Space and Astronautical Science, Japan Aerospace Exploration Agency, 3-1-1 Yoshinodai, chuo-ku, Sagamihara, Kanagawa 252-5210, Japan
[6] Department of Physics & Astronomy, University of California Los Angeles, Los Angeles, CA 90095-1547, USA

## 2. Observations and data reduction

Kepler monitors a ~115 square degree region of sky with ~0.1% photometric errors for a V~15 source, sampling every 30 minutes with a >90%, duty cycle, yielding light curves spanning multiple years for the best-observed sources. Kepler has been used to measure PSDs for Seyfert 1 galaxies (Mushotzky et al. 2011, Carini & Ryle 2012) but, to our knowledge, has not observed any bona fide, highly-variable BL Lacs.

In order to identify targets for Kepler monitoring, Edelson and Malkan (2012) combined WISE, 2MASS and Rosat all-sky survey data to find new AGN. That paper identified W2R1926+42 (RA = $19^h26^m31^s.05$, Dec = $+42°09'59".0$) as a BL Lac at z = 0.155, based on two absorption lines consistent with FeI5269 and NaD5892. Figure 1 shows the historical spectral energy distribution, assembled from the HEASARC, NED, IRSA and MAST archives. The synchrotron peak of the SED lies below $\log(\nu)=10^{14.5}$, consistent with it being a low-frequency peaked blazar (an "LBL," e.g., Padovani & Giommi 1995). The corresponding isotropic monochromatic luminosity is $\nu L_\nu(R)\sim3.6\times10^{44}$ erg/s. The associated radio source, CRATES J192631+420959, is a ~100 mJy flat-spectrum source with ~2%, polarization (Jackson et al. 2007). It is only weakly detected in the X-rays, e.g. ~12 photons in the Rosat all-sky survey (Voges et al. 1999) corresponding to a 0.1-2 keV flux of $\sim5\times10^{-13}$ erg/cm$^2$/s (assuming $N_H = 5\times10^{20}$ cm$^{-2}$ and $\Gamma = 1.7$). It is currently undetected by Fermi.

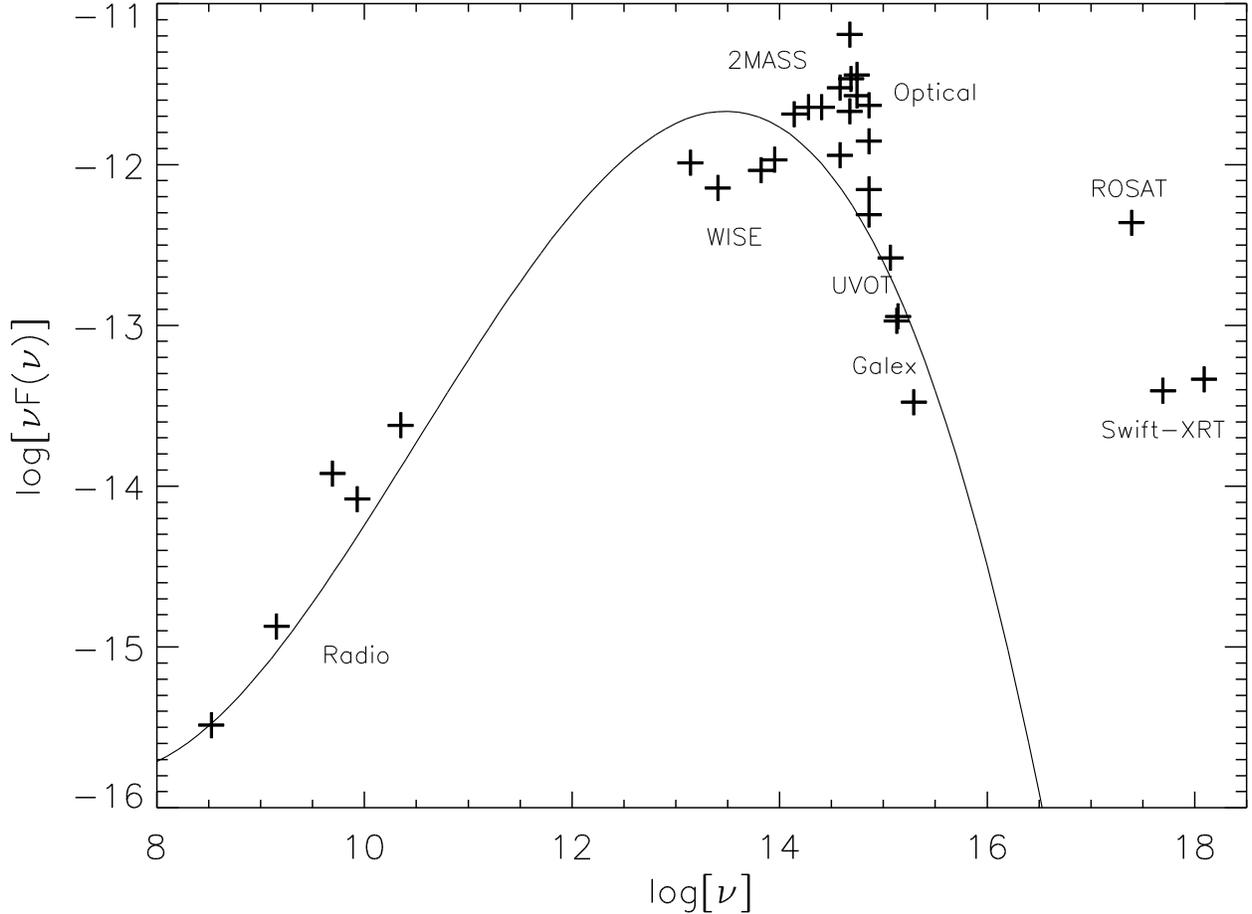

**Figure 1.** SED of the Kepler BL Lac W2R 1926+42. The SED was compiled from non-simultaneous publicly available archival data, and the optical and ultraviolet points have been corrected for Galactic absorption. All points except the X-rays are used to construct the third-order polynomial fit shown in the plot. The synchrotron peak of the SED lies below $\log(\nu)=10^{14.5}$, indicating that this source is an LBL.

Kepler monitoring began in Quarter 11 (Q11).  This paper reports Kepler data for Q11 and Q12, covering 2011 September 29 - 2012 March 28.  The automated Kepler data processing pipeline (e.g., Jenkins et al. 2010) applied four-pixel so-called "optimal" extraction apertures to produce calibrated "SAP_FLUX" light curve files.  The two quarters contained an initial total of 8525 30-minute cadences (samples) over 181 days.

Kepler data can suffer significantly from known problems tracked in the SAP_QUALITY flag, including loss of fine point, focus changes after slews, and data quality degradation due to Solar coronal mass ejection (CME) events.  We eliminated all cadences with any of the SAP_QUALITY bits numbered 2 (safe mode), 3 (Earth point), 4 (coarse point) or 9 (manual exclude, indicating a CME event) set to 1, and the few cadences with missing SAP_FLUX or SAP_FLUX_ERROR data.  This eliminated 502 cadences, ~6% of the original total.  The remaining SAP_QUALITY bits (indicating, e.g., reaction wheel desaturation, "Argabrightening" events, cosmic rays) were not utilized, as these events do not appear to significantly affect the data quality relative to the strong intrinsic variability seen in this source.  The remaining 8023 good cadences yield a net effective duty cycle of ~91%.

In addition Kepler must perform quarterly rolls to keep the Solar panels oriented towards the Sun.  This moves the source to a different chip with a different extraction aperture, inducing small spurious flux offsets between quarters.  To account for this effect we scaled the entire Q12 light curve by a factor of 0.8841 to match the count rate in the last cadence of Q11 to that in the first good cadence of Q12.

The top panel of Figure 2 presents the resulting Kepler Q11/12 SAP_FLUX light curve for W2R1926+42.  This V~17 source has average count rate errors of ~0.2%.  The bottom shows the simultaneous Kepler light curve of W2R1931+43, a newly-discovered Seyfert 1 that falls on the same Kepler module as W2R1926+42.  The W2R1931+43 Q12 light curve was scaled relative to Q11 in the same fashion as W2R1926+42 and then the entire light curve was multiplied by 0.37 so the fractional variability scale is the same for both objects over the range 900-1000 c/s shown in the figure.

Figure 2 demonstrates the differing character of the optical variability of BL Lacs and Seyfert 1s.  The W2R1926+42 light curve gives the impression of repeated strong, rapid flaring, in marked contrast to the smoother, smaller and slower variability seen in W2R1931+43 (and other Seyfert 1 galaxies, e.g., Mushotzky et al. 2011).

The W2R1931+43 light curve also reveals that SAP_FLUX light curves contain uncorrected systematic errors, e.g., the ~1% excursions near Day 126 and Days 154-156.  These are thought due to thermally-induced changes in focus (see Jenkins et al. 2010 for a more detailed discussion) after the satellite returns from safe mode (the gaps on Days 125 and 153).  Because these features are dwarfed by much larger (>10%) variations over timescales of hours in W2R1926+42, no attempt was made to correct for or remove these systematic errors.

In order to allow detailed examination of this rich light curve, in Figures 3 and 4 we show it broken up into eighteen ~10 day segments. These plots show that the flares appear coherent over multiple cadences, with significant variations on timescales of 1-2 hr and frequent >25% (peak-to-peak) variations within a day.  However the apparent flares also appear to be interspersed with longer (often ~2-4 day) periods of relative quiescence.  The ratio of maximum/minimum flux over this 181 day period is 1.79.

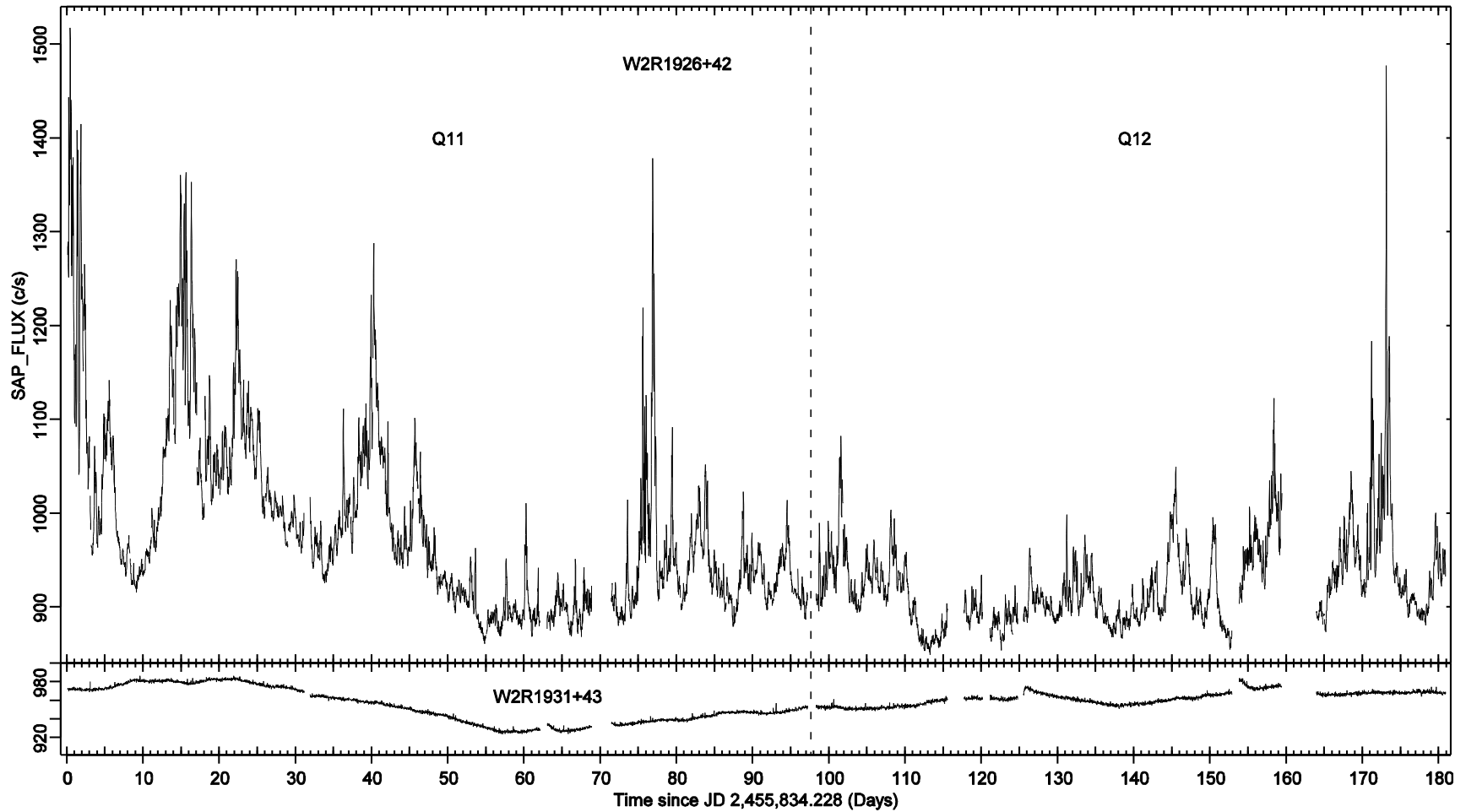

**Figure 2.** Top panel: Kepler light curve of the BL Lac object W2R1926+42, containing 8023 good cadences over 181 days, once every ~30 min with >90% duty cycle and ~0.2% errors. The Q11/Q12 boundary is shown as a dashed line. Bottom panel: comparison light curve of the Seyfert 1 galaxy W2R1931+43, which happens to fall on the same Kepler module as W2R1926+42. The comparison source count rate is scaled so an excursion from 900 to 1000 c/s represents the same percentage change in both plots. The comparison source shows relatively gentle variability typical of Seyfert 1 galaxies, while the target BL Lac object shows much more powerful and erratic variability, with strong, undersampled flaring (e.g. at the start of observations) and long quiescent periods (e.g. near the end) are interspersed throughout. Note also that systematic errors are visible in the comparison light curve (e.g., on Days 154-156 and around Day 126) but these are ~1% effects. The W2R1926+42 light curve was not corrected for these systematics as they are clearly negligible compared to the much larger intrinsic variability.

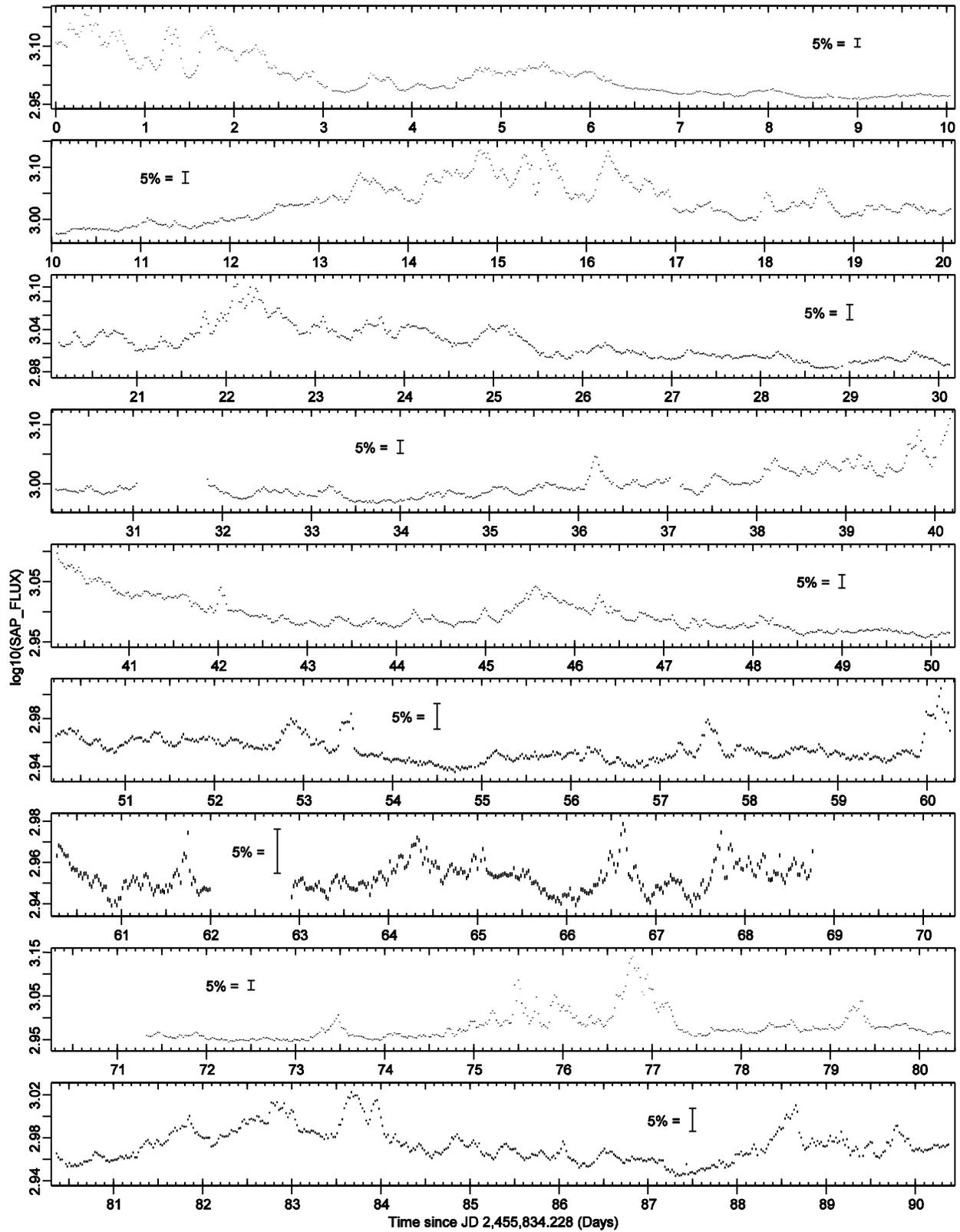

**Figure 3:** Logarithm of first half of the data from the top panel of Figure 2 (Day 0-90.4), split into nine ~10 day segments. In each segment, a 5% bar shows the vertical scale.

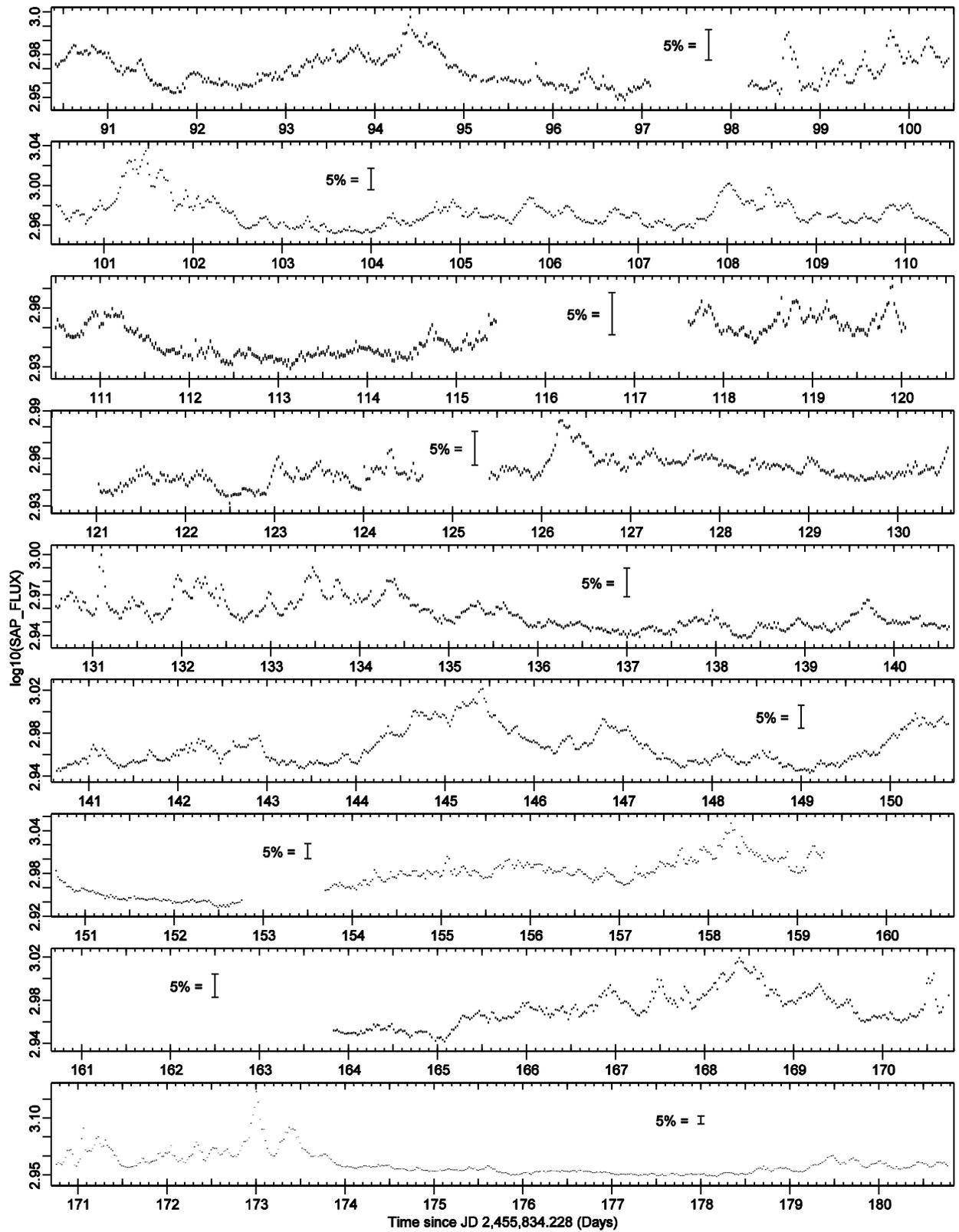

**Figure 4:** Same as Figure 3, except for the second half of the data in Figure 2 (Day 90.4-180.8).

## 3. Time series analyses

While blazars (especially LBLs) can show strong variations on hour timescales, the nature of these variations has not previously been well-measured. This unprecedented Kepler monitoring of a highly-variable blazar allows time-series analyses that could not have been accomplished with any previous dataset, as presented below.

### 3.1. Flux distribution and transformation

Figure 5a shows a histogram of the "raw" measured fluxes for the 8023 30-minute cadences in Figure 1. Two features are apparent. First, there is a clear "floor" in the flux distribution at ~840 c/s (~88% of the mean flux). This could be due to starlight from the underlying galaxy, as well as possible non- or slowly variable components of the central engine/jet. Kepler has large pixels (3.98" across) and a large (4 pixel) aperture was used to extract the light curve. In principle the non-varying stellar component could be estimated and subtracted via direct imaging with HST (e.g., Bentz et al. 2009), but no such observations have yet been performed on W2R1926+42.

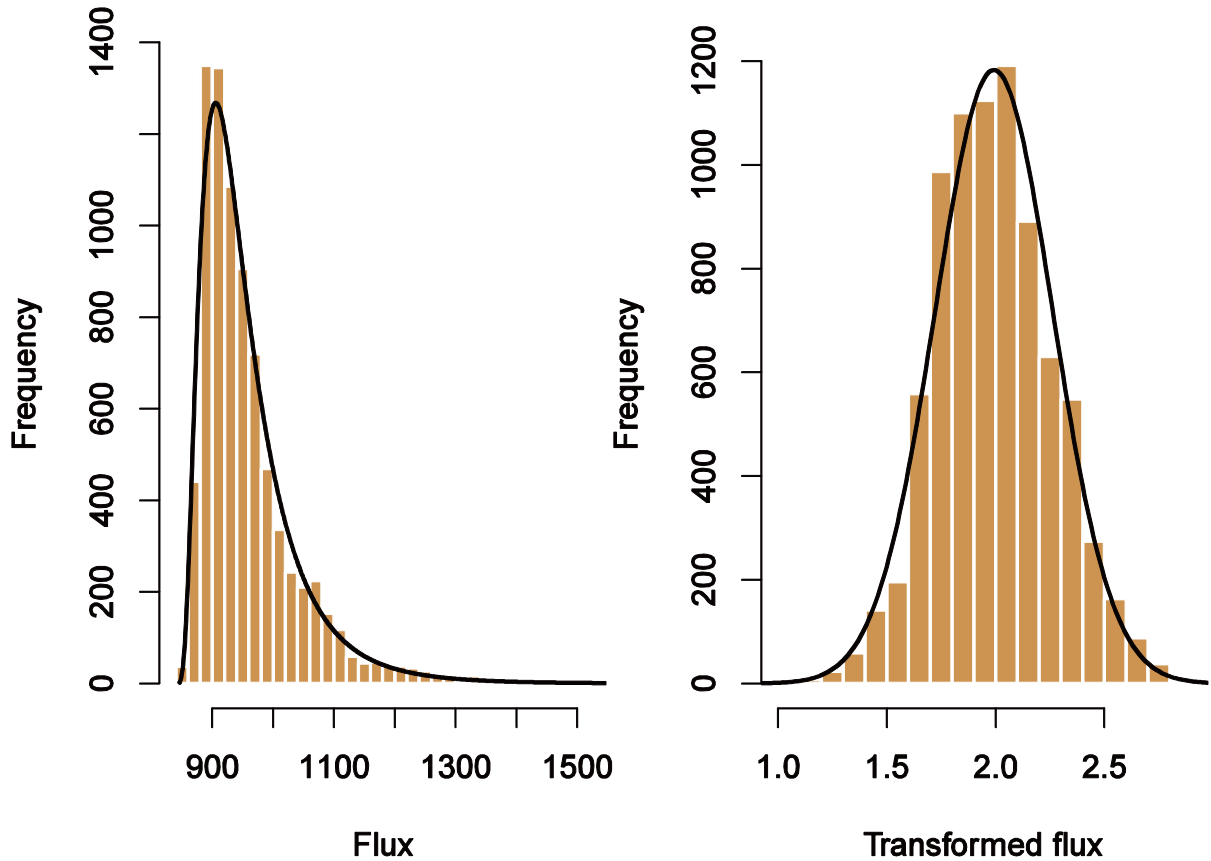

**Figure 5a** (left). Histogram of fluxes from Figure 1. The distribution is highly skewed with a heavy tail to higher fluxes and an apparent "floor" at ~839 c/s. The black curve is the fitted lognormal distribution.
**Figure 5b** (right). Histogram of transformed fluxes. This distribution is closer to a Gaussian, which is plotted as a black curve.

The distribution of fluxes is clearly non-Gaussian, with a strong tail at high fluxes (positive sample skewness, g=2.1; Press et al. 1992). We fitted a lognormal distribution (with a non-zero offset parameter) to this histogram. This has been suggested for X-ray variability of BL Lacs (Giebels & Degrange 2009) and the flux distribution for X-ray binaries is well described by this form (Uttley et al. 2005; Gandhi 2009). The curve (Figure 5a) matches the data reasonably well, although the fit is rather poor. The best fitting offset is $839.0 \pm 0.9$ ct/s. Figure 5b shows the data transformed using $F' = \log_{10}(F-O)$ where F and F' are the original and transformed SAP_FLUX data and O is the flux offset compared to a corresponding Gaussian. The transformed fluxes have almost zero sample skewness (g=0.17).

### 3.2. RMS-flux relation

Seyfert 1 X-ray variations have long been known to exhibit an "rms-flux" relation, wherein the variability amplitude is systematically larger at higher flux levels (e.g., Lyutyi & Oknyanskii 1987, Uttley et al. 2005). This has been less well studied in blazars: an rms-flux relation has been claimed for X-ray variations from BL Lac (Giebels & Degrange 2009), but optical variations of S5 0716+714 (Mocanu & Marcu 2012) gave inconclusive results

We estimated the mean and rms for 655 segments, each 12 cadences long (~0.25 days). Segments containing <10 cadences were ignored. The rms was estimated by subtracting the contribution due to flux errors (Vaughan et al. 2003) and then averaging in flux bins such that each contained ≥50 rms estimates. This yielded 13 rms/flux bins. The resulting plot (Figure 6) shows a clear correlation between rms and flux.

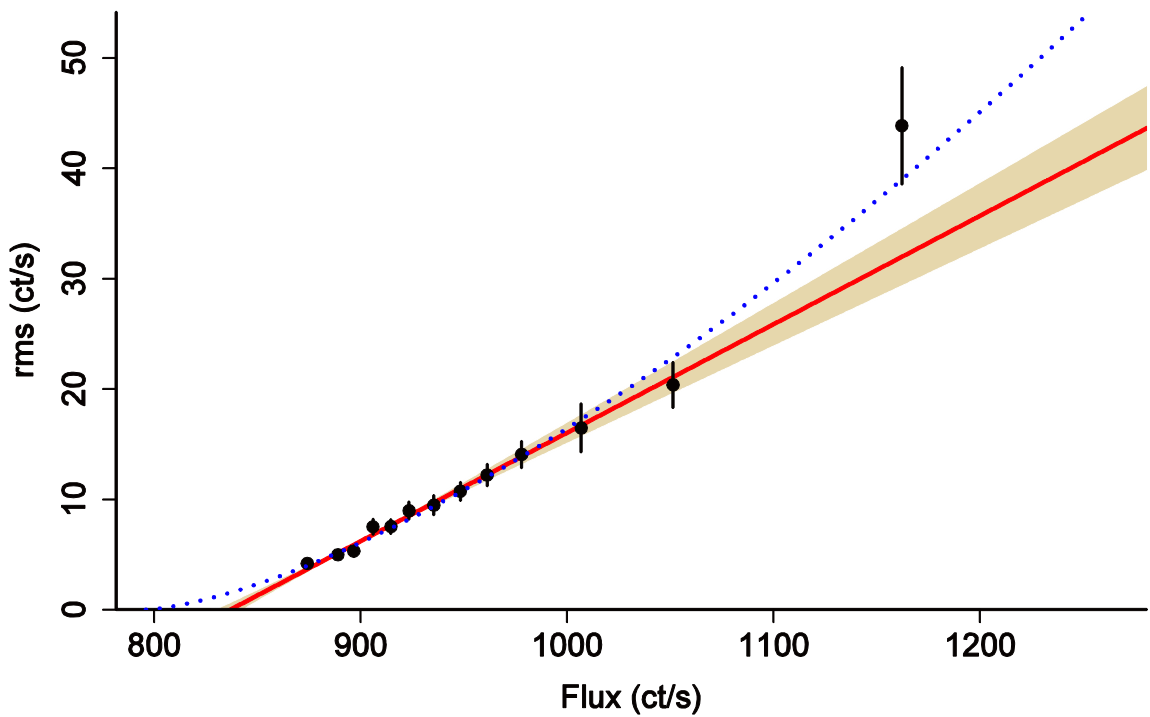

**Figure 6.** RMS-flux relation. The open circles are the average of at least 50 rms estimates calculated from 0.25 day segments. The solid blue line shows the best fitting linear relation (the shaded region shows the corresponding 68.3% confidence band) and the dotted red curve shows the best fitting power-law model.

Fitting these data with a linear model of the form rms=A*(F-O) gave a good fit ($\Delta\chi^2$=11.87 for 11 degrees of freedom, dof) with a flux offset of O = 837 ct/s. This is almost identical to the offset found by fitting the lognormal flux distribution above. A power-law model rms=A*(F-O)$^B$ gave a best fit with an index B = 1.5 ± 0.3. This slightly improved the fit ($\Delta\chi^2$=7.81, 10 dof, corresponding to p=0.04 in a Likelihood Ratio Test), hinting at a divergence from a linear rms-flux relation at the highest fluxes.

### 3.3. PSD analysis

Kepler yields nearly evenly sampled data with the exception of the gaps discussed earlier. In order to estimate the power spectrum on at high frequencies we split the data into M=27 non-overlapping segments of length N=266. The segments were chosen so they include no gaps longer than 10 cadences. The natural choice of time bin size for Kepler data is one cadence, of length dT=1765.46 s. For this and subsequent analyses, all cadences were assumed to have exactly this width. The remaining gaps were filled by linear interpolation (1.5% of data points). The averaged periodogram was averaged in geometrically spaced frequency bins, and errors were computed using standard theory (e.g. van der Klis 1989; Heil, Vaughan & Uttley 2012, Appendix A). The result of the periodogram averaging, shown in Figure 7a, should be a power spectrum estimate with approximately Gaussian errors.

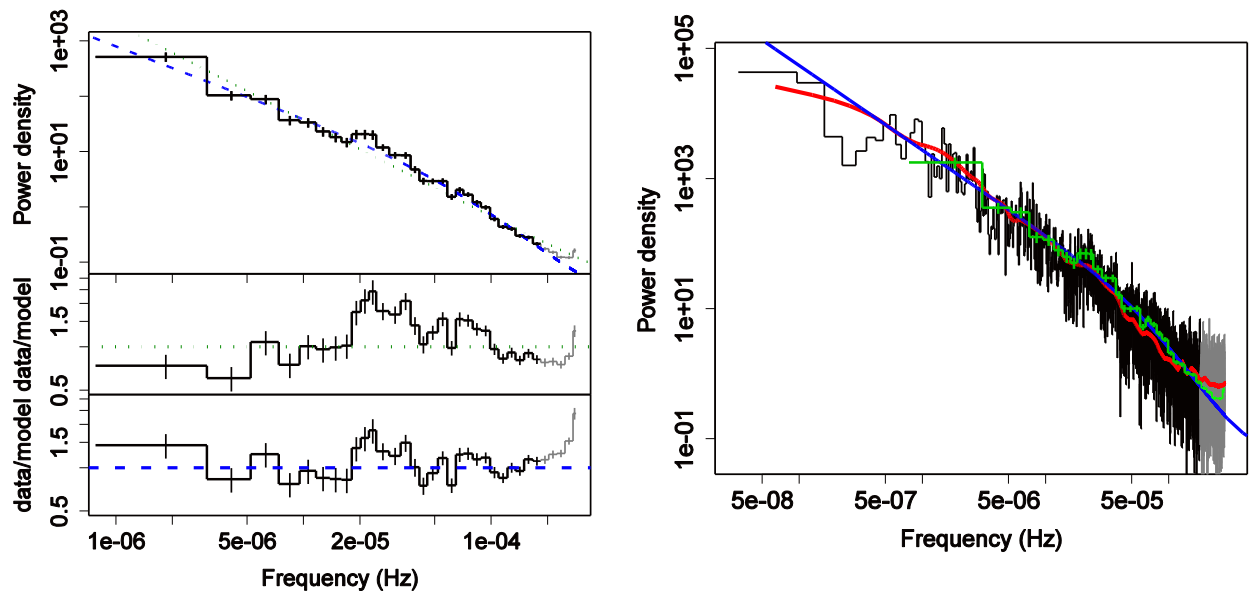

**Figure 7a** (left). Standard method periodogram fits. Top panel shows the periodogram in black, the simple power-law fit in red (dotted) and the bending power-law in green (dashed). The second and third panels show the residuals relative to the simple and bending power-law fits, respectively.

**Figure 7b** (right). ACF-based periodogram (black). A smoothed version (using a 3% LOWESS window) is shown in blue. For comparison, the scaled standard method periodogram is shown in green, and the fit to the standard method periodogram in red.

We fitted simple analytical models to these data using the minimum $\chi^2$ method (using XSPEC; Arnaud et al. 1996). The models were: (a) a simple power-law plus a constant noise term fixed at the level expected from the pipeline flux errors (Vaughan et al. 2003), and (b) a bending power-law plus fixed constant model. These are based on models commonly used to fit Seyfert 1 PSDs (e.g. McHardy et al. 2004). In both cases the model significantly underestimated the power at highest frequencies, so we repeated the fitting including and excluding frequencies above $1.8\times10^{-4}$ Hz (timescales less than 3 cadences). The best-fitting parameters are listed in Table 1. Note that in both cases (including/excluding highest frequencies) the best-fitting bend frequency corresponds to a timescale ~4 hr.

## Table 1: Periodogram fitting results

| Power-law model | High frequencies | $\alpha_1$ | $\alpha_2$ | $\nu_B$ | $\chi^2$ | dof | p |
|---|---|---|---|---|---|---|---|
| Simple | Included | $-1.79 \pm 0.02$ | | | 192.3 | 32 | 8.76E-25 |
| Bending | Included | $-1.07 \pm 0.22$ | $-2.27 \pm 0.12$ | $2.39\text{E-}5 \pm 1.36\text{E-}5$ | 109.1 | 30 | 6.52E-11 |
| Simple | Excluded | $-1.74 \pm 0.02$ | | | 147.6 | 26 | 5.82E-19 |
| Bending | Excluded | $-1.30 \pm 0.14$ | $-2.86 \pm 0.34$ | $6.73\text{E-}5 \pm 3.00\text{E-}5$ | 71.1 | 24 | 1.50E-06 |

**Notes.** Column 1 indicates the type of power-law model fitted and Column 2 shows if data above $1.8 \times 10^{-4}$ Hz were included or excluded in the fit. Models were of the form $P(\nu) = A\nu^{\alpha_1}$ for the simple power-law fits and $P(\nu) = A\nu^{\alpha_2}/(1+(\nu/\nu_B)^{(\alpha_1-\alpha_2)})$ for the bending power-law fits, where $P(\nu)$ is the power at temporal frequency $\nu$, $\nu_B$ is the bending frequency, A is the normalization and $\alpha_1$ and $\alpha_2$ are the low- and high-frequency power-law slopes, respectively. (For the simple power-law fits, $\alpha_1$ gives the slope.) Column 3 gives $\alpha_1$, Column 4 gives $\alpha_2$ and Column 5 gives $\nu_B$, all as defined above. Column 6 gives the measured $\chi^2$ for the fit, Column 7 gives the degrees of freedom in the fit, and Column 8 gives the acceptance probability for the reported values of $\chi^2$ and dof. Note that none of the fits are statistically acceptable, although the bending power-law model in which the high frequencies are excluded is the least unacceptable of the group.

Excluding the high-frequency data gave better fits (see Figure 7a middle and bottom panels), but none gave a statistically acceptable fit (i.e. min $\chi^2$). This is unlike the reasonably well-studied X-ray power spectra of Seyfert 1 galaxies (e.g. Markowitz et al. 2003; McHardy et al. 2004) which are well described by a similar model. However, the above analysis does not explicitly include the effects of spectral leakage (Uttley et al. 2002), which may distort the spectrum, particularly at highest frequencies. But even in the absence of such effects it should be noted that the present dataset is of considerably higher quality than obtained for Seyfert galaxies: having many more data points with smaller flux errors and the source itself is more highly variable. The PSD estimate is therefore more sensitive to small deviations around a simple power law-based model and thus it is no surprise that the quality of the fit if poorer.

The segment averaging used above and gives a spectral estimate with approximately Gaussian errors, but at expense of ignoring the lowest frequencies. In order to examine the lowest frequencies contained in the full 181 day light curve we briefly explored a novel method of spectral estimation. We first estimated the autocorrelation function (ACF) using the Discrete Correlation Function (DCF; Edelson & Krolik 1988), and then Fourier transformed this to yield a periodogram. The DCF was designed to measure correlation functions from unevenly sampled data such as these Kepler light curves. It essentially averages the contributions to ACF($\tau$) from data pairs whose time difference lies in the corresponding lag bin; the only effect of gaps is to diminish the number of such contributions in the corresponding bins. Because it does not break the time series into segments, it can access lower temporal frequencies. The result is shown in Figure 7b. As the statistical properties of this estimate are not yet well understood, we did not fit these data. Nevertheless, Figure 7b shows this estimate is consistent the standard estimate, and with the bending power-law model discussed above.

We also searched for narrow QPOs using the methods outlined in Vaughan (2010). This revealed no strong candidates, but the fit residuals (lower panels of Figure 7) do show a broad excess near $\sim 2 \times 10^{-5}$ Hz. In principle this unexplained structure could due to a very broad (Q~2-3) QPO, a second bend/break in PSD, an artifact due to non-stationarity, or it could be another indication of the limits of PSD analysis.

## 4. Discussion

Previous "microvariabilty" studies could only obtain continuous ~12 hr light curves, and thus were focused on determining if the variations seen in a single night were discrete events or part of longer duration changes (e.g., Ghosh et al. 2001). These data, nearly continuous over ~6 months, show incessant, rapid flaring and a bending power-law PSD. Further, the flare amplitudes are not normally distributed, but instead are skewed towards high fluxes.

The standard model holds that broadband blazar variability (e.g., Blandford & Koenigl 1979) originates in shocks within the jet. Relativistic time dilation causes characteristic timescales to be compressed by the Doppler factor γ, typically ~10. Correcting for the beaming factor, the observed PSD break at ~4 hr corresponds to a rest-frame variability timescale of ~1-2 days, much less than the >1 month optical break timescales in Seyfert 1s (Mushotzky et al. 2011, Carini & Ryle 2012). Because the standard model holds that optically-thin synchrotron radiation from LBLs falls within the Kepler band, this light curve may be the first to probe the inner jet of a blazar with such extraordinary sampling and sensitivity.

The highly-skewed flux distribution and monotonic increasing relation between rms and flux arise naturally in the Biteau & Giebels (2012) "minijets-in-a-jet" model, in which a smaller region within a relativistic jet (e.g. a shock) produces "minijets" with random orientations in their rest frame. In this context, the effective relativistic Lorentz factor of the emitting region is the product of the Lorentz factors of both the jet and minijet (e.g. Γ~100 for Mrk 501 and PKS 2155-304; Giannios, Uzdensky & Begelman 2009). The observed bending timescale would correspond to a boosting-corrected timescale of ~400 hours, or ~0.5 months, in this scenario. This begins to approach the break timescales seen in Seyferts, The characteristic size of the emission region where this variability originates must be compact, even if Lorentz factors are as high as those seen in Mrk 501. Assuming a typical black hole mass of $~10^8$—$10^9$ $M_{sun}$, the emission zone must be smaller than ~1000—100 Schwarzschild radii. Alternatively, similar skewed flux distributions and rms-flux relations are also seen in X-ray variability of Seyfert 1s (e.g., Uttley et al. 2005) and Galactic X-ray binaries (e.g., Uttley & McHardy 2001) and optical variability of X-ray binaries (Gandhi 2009) and at least one CV (Scaringi et al. 2012). Their fluctuations have been successfully modeled as multiplicative perturbations in an accretion disk (e.g. Lyubarskii 1997). If the blazar jet threads the disk, then one might expect some relation between disk perturbations and jet emission. Such a "disk-jet connection" has been discussed on a statistical basis in ensemble studies (e.g. Maraschi & Tavecchio 2003).

The isotropic luminosities of blazars and Seyfert 1s are roughly comparable while their rest frame bending timescales appear to differ by at least an order of magnitude, suggesting that different mechanisms are responsible for optical variability in these two classes of AGN. We also note that Ries et al. (2012) found that the beamed transient SWIFT1644 shows a similar PSD with a break in the few mHz range and similar slopes at both high and low frequencies. We speculate that this could be a characteristic property of beamed sources observed with sufficiently high signal-to-noise and dense sampling.

The periodograms produced by the discrete autocorrelation function and standard methods appear consistent. However we were unable to obtain a statistically acceptable fit with any simple 4 free parameter model. The residuals to the power law fits show coherent detail which is not well modeled by the simple curving power law PSD models and require more fitting parameters. This indicates that the combination of Kepler's sensitivity and sampling and this blazars' strong variability is, for the first time, pushing us clearly beyond the limits of conventional PSD model-fitting analysis.

## 5. Conclusions

Kepler light curves are ideally suited to studying blazars' rapid optical variability. These observations of the newly-discovered Kepler BL Lac W2R1926+42 open a new window on blazar optical variability, revealing what was once called "microvariability" to be a combination of strong flaring and quiescent periods. The source shows a strong rms/flux relation with a floor of about 88% of the mean flux level. Our extensive PSD modeling was unable to produce a fully satisfactory fit with simple (4 parameter) models, although the most acceptable results indicate a downwardly curving power-law with a characteristic ~4 hr timescale.

Our picture of blazar variability will continue to improve as Kepler continues to monitor this blazar (and hopefully others). Short cadence (1 minute) sampling of W2R1926+42 has already begun, possibly providing further key constraints. The observed light curve segments in Figures 3 and 4 show significant variations on ~1 hr timescales, suggesting that flaring activity on timescales faster than the shortest time sampling of 0.5 hr may remain to be uncovered. These data will also allow detailed study and testing of the apparent excess variability at short timescales. Likewise the extension of this light curve from ~6 months to many years as the satellite continues to gather data will allow application of even more sophisticated analysis techniques. Kepler's unprecedented sampling and precision will require application of both time-series analytical tools and theoretical models that can describe the new features of the time series of these of relativistically-beamed AGN.

**Acknowledgements:** The authors appreciate the helpful assistance of the Kepler GO office in scheduling and understanding these observations, as well as the editor and anonymous referee for a timely and useful review. This research utilized the HEASARC, IRSA, NED and MAST data archives and the NASA Astrophysics Data System Bibliographic Service. RE and RM acknowledge support by the Kepler GO program through NASA grants NNX11AC81G, NNX12AC93G, and NNX13AC26G. JS acknowledges Joe Bredekamp and the NASA Applied Information Systems Research Program for support.